\documentclass[aps,prl,
 amsmath,amssymb,
 twocolumn, groupedaddress,
 showpacs]{revtex4-1}
\usepackage{bm}
\usepackage{graphicx}
\usepackage{gensymb}
\usepackage{amsmath}
\usepackage{amssymb}
\usepackage{ulem}
\usepackage{here}

\begin{document}

\title{Observation of Chiral Fermions with a Large Topological Charge and Associated Fermi-Arc Surface States in CoSi}

\author{Daichi Takane,$^1$ Zhiwei Wang,$^2$ Seigo Souma,$^{3,4}$ Kosuke Nakayama,$^1$ Takechika Nakamura,$^1$ Hikaru Oinuma,$^1$ Yuki Nakata,$^1$ Hideaki Iwasawa,$^5$ Cephise Cacho,$^5$ Timur Kim,$^5$ Kouji Horiba,$^6$ Hiroshi Kumigashira,$^{6,7}$ Takashi Takahashi,$^{1,3,4}$ Yoichi Ando,$^2$ and Takafumi Sato$^{1,3}$}

\affiliation{$^1$Department of Physics, Tohoku University Sendai 980-8578, Japan\\
$^2$Physics Institute II, University of Cologne, 50937 K\"oln, Germany\\
$^3$Center for Spintronics Research Network, Tohoku University, Sendai 980-8577, Japan\\
$^4$WPI Research Center, Advanced Institute for Materials Research, Tohoku University, Sendai 980-8577, Japan\\
$^5$Diamond Light Source, Harwell Science and Innovation Campus, Didcot, Oxfordshire OX11 0QX, UK\\
$^6$Institute of Materials Structure Science, High Energy Accelerator Research Organization (KEK), Tsukuba, Ibaraki 305-0801, Japan\\
$^7$Institute of Multidisciplinary Research for Advanced Materials (IMRAM), Tohoku University, Sendai 980-8577, Japan
}

\date{\today}

\begin{abstract}
Topological semimetals materialize a new state of quantum matter where massless fermions protected by a specific crystal symmetry host exotic quantum phenomena. Distinct from well-known Dirac and Weyl fermions, structurally-chiral topological semimetals are predicted to host new types of massless fermions characterized by a large topological charge, whereas such exotic fermions are yet to be experimentally established. Here, by using angle-resolved photoemission spectroscopy, we experimentally demonstrate that a transition-metal silicide CoSi hosts two types of chiral topological fermions, spin-1 chiral fermion and double Weyl fermion, in the center and corner of the bulk Brillouin zone, respectively. Intriguingly, we found that the bulk Fermi surfaces are purely composed of the energy bands related to these fermions. We also find the surface states connecting the Fermi surfaces associated with these fermions, suggesting the existence of the predicted Fermi-arc surface states.  Our result provides the first experimental evidence for the chiral topological fermions beyond Dirac and Weyl fermions in condensed-matter systems, and paves the pathway toward realizing exotic electronic properties associated with unconventional chiral fermions.
\end{abstract}


\maketitle
The search for new fermionic particles, which has long been an important research target in elementary-particle physics, is now becoming an exciting challenge in condensed-matter physics to realize exotic fermions, as highlighted by the discovery of two-dimensional (2D) Dirac fermions in graphene \cite{CastroNetoRMP2009} and helical Dirac fermions at the surface of three-dimensional (3D) topological insulators \cite{HasanReview, SCZhangReview, AndoReview}. Recent discovery of 3D Dirac semimetals and Weyl semimetals hosting massless Dirac/Weyl fermions \cite{WangPRB2012, WangPRB2013, NeupaneNC2014, BorisenkoPRL2014, LiuScience2014, XuScience2015, LvPRX2015, YangNP2015, SoumaPRB2016, ArmitageRMP2018} further enriches the category of exotic fermions. Majorana fermions in topological superconductors, realized intrinsically \cite{SatoRPP2017} or artificially \cite{MurikScience2012, DasNP2012, Nadj-PergeScience2014, DengScience2016}, are an attractive target for the quest of novel fermions due to its potential applicability to fault-tolerant quantum computation.

 While all these fermions could manifest themselves in elementary-particle physics, recent theories have predicted new types of massless fermions in condensed-matter systems which have no counterparts in elementary particles \cite{JLManesPRB2012, BBradlynScience2016, HWengPRB2016, ZZhuPRX2016, BLvNaure2017}. Such fermions appear in the crystals with specific space groups, as exemplified by the spin-1 chiral fermion and the double Weyl fermion showing multifold band degeneracies at the band-crossing point (node) protected by the crystal symmetry \cite{BBradlynScience2016, PTangPRL2017, GChangPRL2017}. As shown in Fig. 1(a), a well-known spin-1/2 Weyl fermion shows a Weyl-cone energy band dispersion with two-fold degeneracy at the node that carries the topological charge (Chern number) $C$ of $\pm$1 \cite{BBradlynScience2016, PTangPRL2017, GChangPRL2017}. On the other hand, the spin-1 chiral fermion is characterized by the combination of a Dirac-like band and a flat band with three-fold band crossings, carrying a larger topological charge of $\pm$2 \cite{JLManesPRB2012, BBradlynScience2016, PTangPRL2017}. Double Weyl fermion, which appears under the time-reversal-invariant condition, exhibits four-fold degeneracy at the node that also carries topological charge of $\pm$2 \cite{BBradlynScience2016, PTangPRL2017, YXuPRA2016}. Such fermions provide an excellent platform for the observation of unconventional bulk chiral transport and exotic circular photogalvanic effects \cite{LBalentsPhys2011, TOjanenPRB2013, JMaPRB2015, SZhongPRL2016, CFangNP2016, QMaNP2017, FdJuanNC2017, GChangPRL2017, FFlickerarXiv2018}.

\begin{figure}
\includegraphics[width=2.8in]{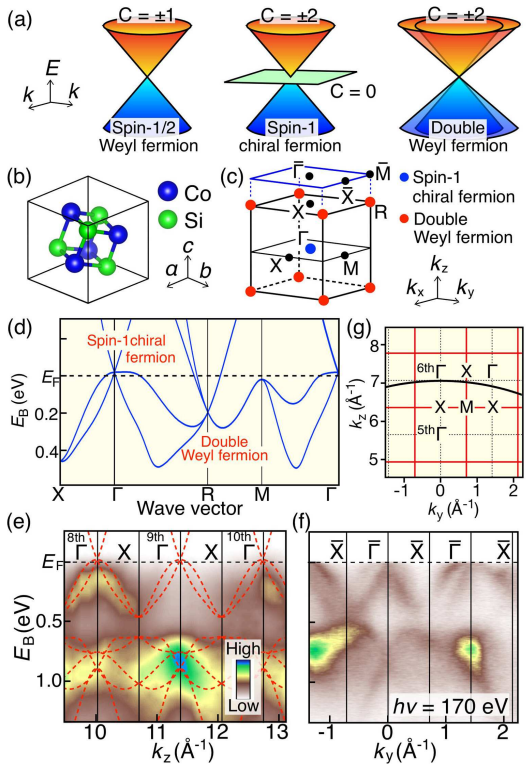}
\caption{(Color online) (a) Schematic band dispersions in 3D $E$-$k$ space for the spin-1/2 Weyl fermion (left), spin-1 chiral fermion (middle), and double Weyl fermion (right). (b), (c) Crystal structure and bulk/surface BZs of CoSi, respectively. Blue and red dots in (c) represent the $k$ points where the band-crossing points exist for spin-1 chiral fermion and double Weyl fermion, respectively. (d) Calculated bulk-band structure near $E_{\rm F}$ of CoSi along high-symmetry cuts in the bulk BZ obtained by our first-principles band-structure calculations. Note that the calculated band structure is essentially the same as that previously reported \cite{PTangPRL2017}. (e) Plot of the normal-emission ARPES intensity as a function of $k_z$ (corresponding to the out-of-plane $\Gamma X$ cut). Inner potential was set to be $V_0$ = 20 eV from the periodicity of the band dispersion. Red dashed curves represent the calculated band structure along the $\Gamma X$ cut. (f) ARPES intensity along the $\bar \Gamma \bar X$ cut measured with $h\nu$ = 170 eV. (g) Bulk BZ in the $k_y$-$k_z$ plane together with measured cut at $h\nu$ = 170 eV. The BZ size is based on the experimental lattice constant of $a$ = 4.438 \AA.}
\end{figure}

 Recently, it was proposed from first-principles band-structure calculations \cite{PTangPRL2017, GChangPRL2017} that cubic crystals of CoSi family including CoGe, RhSi, and RhGe [space group $P$2$_1$3 (No. 198) \cite{PDemchenkoCMA2008}; see Figs. 1(b) and 1(c) for the crystal structure and the bulk Brillouin zone (BZ)] host various types of exotic fermions owing to the chiral crystal structure. As shown in Fig. 1(d), the band calculations without the spin-orbit coupling (SOC) suggests that CoSi hosts two types of symmetry-protected unconventional crossings of linearly dispersive bands, one at the $\Gamma$ point with six-fold degeneracy (including spin) and another at the $R$ point with eight-fold degeneracy, which give rise to spin-1 chiral fermions with $C$ = 2 and double Weyl fermions with $C$ = -2, respectively \cite{PTangPRL2017}. Such chiral fermions also give rise to Fermi-arc surface states traversing the projection of bulk bands associated with these fermions \cite{PTangPRL2017}. It is thus highly desirable to directly observe such characteristic features in experiments (note that the inclusion of SOC in the calculations further causes a small band splitting together with partial lifting of the band degeneracy at $\Gamma$ and $R$).
 
 In this article, we report angle-resolved photoemission spectroscopy (ARPES) of CoSi single crystal. By utilizing energy-tunable photons from synchrotron radiation, we have determined the band structure in the 3D BZ, and found that this material hosts exotic chiral fermions and associated Fermi-arc surface states, in accordance with the theoretical predictions. Moreover, the low-energy excitations were found to be solely dictated by the energy bands related to these fermions, providing an excellent opportunity to realize unconventional transport properties. The present study experimentally establishes the existence of chiral topological fermions beyond Dirac and Weyl fermions in condensed-matter systems.

High-quality single crystals of CoSi were synthesized with a chemical vapor transport method. ARPES measurements were performed at the beamlines in DIAMOND Light Source, Photon Factory, KEK, and UVSOR. First-principles band-structure calculations were carried out using QUANTUM ESPRESSO code with generalized gradient approximation \cite{GiannozziJPhys2015}. For details, see Section 1 of Supplemental Material.
 
 We first discuss the overall 3D electronic structure of CoSi. Figure 1(e) shows the valence-band dispersion along the wave vector perpendicular to the sample surface ($k_z$), measured with the normal-emission setup by varying $h\nu$ in the soft-x-ray region (310-630 eV). One can find some energy bands displaying a prominent $k_z$ dispersion, e.g., in the range of 0 - 0.5 eV and 0.7 - 1.2 eV in terms of the binding energy $E_{\rm B}$. The observed band dispersions well follow the periodicity of the bulk BZ; for example, the near-$E_{\rm F}$ band has a top (bottom) of the dispersion at each $\Gamma$($X$) point, consistent with the bulk-band calculations without SOC (red dashed curves), while the spectral intensity is markedly suppressed around the middle $\Gamma$ point due to the matrix-element effect of the photoelectron intensity. The periodic dispersions along the $k_z$ direction indicate the bulk origin of the observed bands. As shown in Fig. 1(f), we carried out ARPES measurements along the in-plane $\bar \Gamma \bar X$ cut with $h\nu$ = 170 eV which well traces the $\Gamma X$ line around the center of BZ involving the 6th $\Gamma$ point [Fig. 1(g)] (note that the ${\bm k}$  cut gradually deviates from the $\Gamma X$ line on entering the next BZ), and found that the overall dispersive feature is similar to that along the out-of-plane $\Gamma X$ cut [Fig. 1(e)], although fine spectral features, such as the presence of two holelike bands near-$E_{\rm F}$ around $\Gamma$, are better resolved at $h\nu$ = 170 eV due to the improved energy and $k$ resolutions.
 
 \begin{figure}
\includegraphics[width=2.8 in]{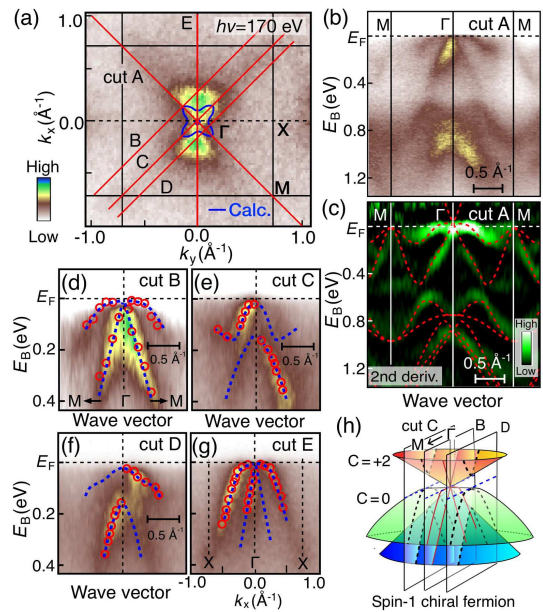}
\caption{(Color online) (a) ARPES-intensity mapping at $E_{\rm F}$ plotted as a function of $k_x$ and $k_y$ measured at $h\nu$ = 170 eV. Intensity at $E_{\rm F}$ was obtained by integrating the spectral intensity within $\pm$10 meV of $E_{\rm F}$. Calculated FS is overlaid by blue curve. (b) ARPES intensity and (c) its second-derivative intensity measured along the cut A in (a) ($\Gamma M$ cut). The second-derivative plot in (c) is based on the EDC divided by the Fermi-Dirac (FD) function at $T$ = 25 K convoluted with the resolution function. Note that this division is useful to avoid appearance of a false flat band at $E_{\rm F}$ originating from the background with a finite FD cut-off. Red dashed curves in (c) represent the calculated band structure along the $\Gamma M$ line. (d)-(g) ARPES intensity measured along the cuts B-E in (a), respectively. Red open circles are the energy position of bands determined by tracing the peak position of the EDCs divided by the FD function at $T$ = 25 K convoluted with the resolution function, while the blue dashed curves are a guide to the eyes to trance the experimental band dispersion obtained by assuming symmetric band dispersions with respect to the $\Gamma M$ line (vertical dashed line). (h) Schematic band dispersion for the spin-1 chiral fermion at $\Gamma$, together with the measured cuts (cuts B-D).}
\end{figure}
 
 Now that we have a good control of $k_z$ with respect to the bulk BZ, we next search for the band dispersion characteristic of the predicted spin-1 chiral fermion at the $\Gamma$ point. For this sake, we fix the $h\nu$ value to 170 eV that well traces the $k_z$ = 0 plane ($\Gamma XM$ plane) around the center of BZ and maps out the near-$E_{\rm F}$ band structure as a function of in-plane wave vectors ($k_x$ and $k_y$). The ARPES-intensity mapping at $E_{\rm F}$ in Fig. 2(a) displays a strong intensity around the $\Gamma$ point with a two-fold symmetric pattern elongated toward the $M$ points. Such an anisotropy could be explained in terms of the strong anisotropy of the Fermi surface (FS) itself as seen in the calculation (blue curve) combined with the dipole selection rule of the photoelectron intensity. To gain further insight into the band structure around $\Gamma$, we plot in Figs. 2(b) and 2(c) the ARPES intensity and its second derivative, respectively, measured along the $\Gamma M$ cut [cut A in Fig. 2(a)]. One can clearly resolve two holelike bands centered at the $\Gamma$ point near $E_{\rm F}$; the inner band linearly disperses toward $E_{\rm F}$ on approaching the $\Gamma$ point, while the outer band shows a nearly flat dispersion pinned at $E_{\rm F}$ around the $\Gamma$ point. Intriguingly, the inner and outer bands appear to intersect with each other exactly at the $\Gamma$ point, in overall agreement with the calculated band structure showing an intersection of the flat band and the X-shaped band just at $\Gamma$ [see red dashed curves around $\Gamma$ in Fig. 2(c)]. Our spectral analysis (Section 2 of Supplemental Material) suggests a quantitative mismatch in the band-crossing point at $\Gamma$; it is located slightly ($\sim$ 5 meV) below $E_{\rm F}$ in the experiment, whereas it is 15 meV above $E_{\rm F}$ in the calculations. As shown in Fig. 2(d), the ARPES intensity along another $\Gamma M$ cut (cut B) exhibits an intersection of two bands at $\Gamma$, consistent with the data along cut A. We show in Figs. 2(e) and 2(f) the ARPES intensity along two representative ${\bm k}$ slices which do not pass the $\Gamma$ point [cuts C and D in Fig. 2(a)]. One can recognize that the inner and outer bands do not contact with each other irrespective of ${\bm k}$, supporting that the band degeneracy occurs only at the $\Gamma$ point, which is also confirmed by the ARPES intensity along the $\Gamma X$ cut [cut E in (a)] in Fig. 2(g). Such behavior is schematically shown by a band diagram in Fig. 2(h). All these results suggest the existence of spin-1 chiral fermions in CoSi. Note that the possibility that a small SOC-induced gap may open is not excluded because of the present finite experimental resolution (see Section 2 of Supplemental Material).

 Next, we investigate the double Weyl fermion at the $R$ point. Figure 3(a) displays the ARPES intensity at $E_{\rm F}$ as a function of $k_x$ and $k_y$ measured at $h\nu$ = 136 eV ($k_z$ = $\pi$; the $RMX$ plane). One can recognize the bright intensity at the $R$ point, consistent with the existence of a small electron pocket in the calculation (blue circle) which originates from the upper branch of double Weyl-fermion dispersion [Fig. 1(d)], whereas the size of FS appears to be smaller in the experiment. To visualize the band dispersion associated with the double Weyl fermion, we have performed high-resolution ARPES measurements along the ${\bm k}$ cut which crosses the $R$ point [cut A in Fig. 3(a)]. As seen in the ARPES intensity and its second derivative plots in Fig. 3(b), there exists an X-shaped band which is degenerate at the $R$ point at $E_{\rm B} \sim$ 0.1 eV. This dispersion is qualitatively reproduced by the calculation (left panel), while the experimental nodal point is closer to $E_{\rm F}$ than the calculation by $\sim$ 0.1 eV. It is predicted from the calculations that the two Weyl cones always keep the band degeneracy on the whole $RMX$ ($k_z$ = $\pi$) plane due to the crystal symmetry \cite{PTangPRL2017}, so that the splitting of the double Weyl cones should not be visible in Fig. 3(b). To verify the existence of the double Weyl cone, one must select a ${\bm k}$ cut which crosses the $R$ point by varying $k_z$ \cite{PTangPRL2017}. For this sake, we have carefully traced the band dispersion along the $\Gamma R$ cut by varying simultaneously the photon energy (i.e. $k_z$) and the in-plane wave vector, since the large splitting of the Weyl cones is theoretically predicted in this cut [see Fig. 1(d)]. The second-derivative ARPES intensity in Fig. 3(c) and corresponding EDCs in Fig. 3(d), corroborated by soft-x-ray ARPES data (Fig. S3 of Supplemental Material) suggest that the lower branch of the X-shaped band splits into two branches on approaching $\Gamma$ from $R$. The magnitude of splitting reaches the maximum value of $\sim$ 0.4 eV midway between the $\Gamma$ and $R$ points ($k_z \sim$ 0.5$\pi$) and then gradually decreases on approaching the $\Gamma$ point to form the spin-1-chiral-fermion dispersion, whereas such splitting is absent along the $RM$ cut, in accordance with the calculated band dispersion (red dashed curves). In contrast to the large splitting of the lower Weyl-cone branches along $\Gamma R$, we do not observe a detectable splitting of the upper Weyl-cone branches in Fig. 3(c), consistent with the calculation. These results support the existence of double Weyl fermion in CoSi.
 
 \begin{figure}
\includegraphics[width=3 in]{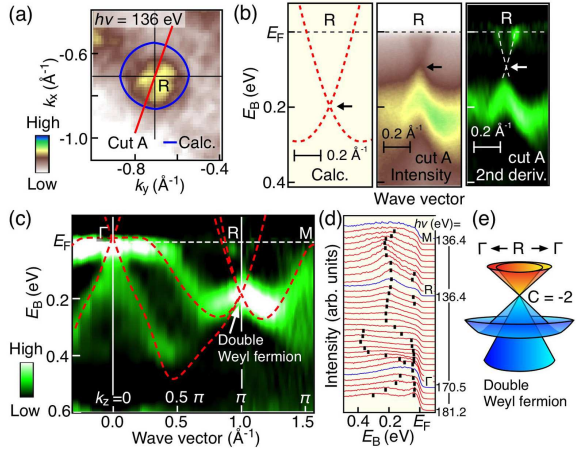}
\caption{(Color online) (a) ARPES-intensity mapping at $E_{\rm F}$ plotted as a function of $k_x$ and $k_y$ around the $R$ point measured at $h\nu$ = 136 eV ($k_z\sim\pi$ plane). Calculated Fermi surface is shown with a blue curve. (b) Calculated band structure (left), ARPES intensity (middle), and the second-derivative of the ARPES intensity (right), measured along cut A in (a) which crosses the $R$ point. ARPES intensity along cut A (parallel to the analyzer slit; tilted from the high-symmetry line due to sample alignment) was obtained by simultaneously collecting fine angular images of photoelectrons. (c) Second-derivative of the ARPES intensity along $\Gamma R$ and $RM$ cuts, together with the calculated band structure (red dashed curves). (d) Corresponding raw EDCs whose peak position is indicated by black dots. (e) Schematic band dispersion for double Weyl fermion at the $R$ point.}
\end{figure}
 
To search for the predicted Fermi-arc surface states, we performed ARPES measurements using vacuum-ultraviolet photons. We selected $h\nu$ = 67 eV which corresponds to $k_z$ = 0.75$\pi$  at $k_x\sim$ 0 [in reduced BZ; see Fig. 4(b)] to avoid a possible complication from the bulk bands. As shown in Fig. 4(a), the ARPES intensity is not localized around $\bar \Gamma$ unlike the case of $h\nu$ = 170 eV [Fig. 2(a)], but apparently elongated toward two of four adjacent $\bar M$ points, producing a C$_2$-symmetric ``Z''-shaped intensity pattern. This feature is commonly observed at other $h\nu$'s [Fig. 4(b)] despite a sizable change in the $k_z$ value [0.3-0.95$\pi$; Fig. 4(b)], suggestive of its surface origin. The ARPES intensity and its second-derivative plot  along cut A [red line of Fig. 4(a)] in Figs. 4(c) and 4(d) signify a dispersive weak feature across $E_{\rm F}$, besides the bulk band topped at $E_{\rm B} \sim$ 0.1 eV. This band has a negative slope and appears only in the positive $k$ region [see also Fig. 4(d)]. Interestingly, as shown in Fig. 4(e), such dispersion relation is reversed in cut B located at the opposite side of cut A with respect to the $\bar \Gamma$ point [Fig. 4(a)]; namely, the band has a positive slope and appears only in the negative $k$ region. As seen by a comparison between Figs. 4(f) and 4(d), the energy dispersion of the $E_{\rm F}$-crossing band seems to be unchanged between $h\nu$ = 61 and 67 eV. In fact, the experimental band dispersion extracted from the second derivative of the momentum distribution curves (MDCs) in Fig. 4(g) signifies that the location of the $E_{\rm F}$-crossing band is robust against the $h\nu$ variation, in contrast to the bulk band. These results suggest the surface origin of the $E_{\rm F}$-crossing band. This band is likely attributed to the predicted Fermi-arc surface states \cite{PTangPRL2017}, since the calculated surface states exhibit the C$_2$-symmetric Fermi contour and connects with the bulk FSs at the $\bar \Gamma$ and $\bar M$ points, consistent with the present ARPES results.
 
\begin{figure}
\includegraphics[width=3 in]{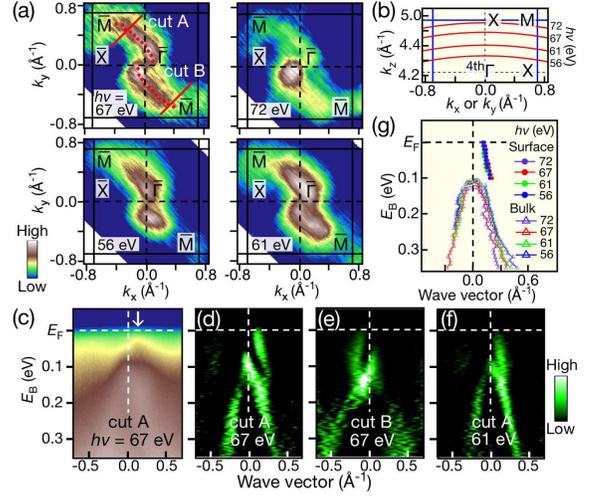}
\caption{(Color online) (a) FS mapping of CoSi at $h\nu$ = 56-72 eV. (b) Measured cuts in $k_x$-$k_z$ ($k_y$-$k_z$) plane. (c), (d) ARPES intensity and second-derivative intensity of MDCs, respectively, measured along cut A in (a) (red line) at $h\nu$ = 67 eV. White arrow in (c) indicates the $E_{\rm F}$-crossing point of surface states. (e) Same as (c) but measured along cut B in (a). (f) Same as (d) but measured at $h\nu$ = 61 eV. (g) Experimental band dispersions extracted from the second derivative of MDCs at various $h\nu$'s.}
\end{figure}
 
 Now we discuss implications of the present ARPES results in relation to the band calculations. While the band dispersion shows a qualitatively good agreement between the experiment and calculations, there exists a finite quantitative difference in the energy position of the nodal points. The calculation predicted that the nodal points for the spin-1 chiral fermion at $\Gamma$ and the double Weyl fermion at $R$ are 15 meV above $E_{\rm F}$ and 0.2 eV below $E_{\rm F}$, respectively [Fig. 1(d)]. On the other hand, in this experiment, they are located at slightly ($\sim$ 5 meV) below $E_{\rm F}$ [Figs. 2(d) and 2(g); Fig. S1 of Supplemental Material] and 0.1 eV below $E_{\rm F}$ [Fig. 3(b)], respectively, suggesting a band-renormalization effect. Importantly, our data verify that the band never crosses $E_{\rm F}$ near the $M$ point (Section 4 of Supplemental Material) as predicted by the calculations, leaving only the novel fermions near $\Gamma$ and $R$ to be relevant to low-energy physics. This would be favorable for observing predicted unconventional physical properties related to the chiral fermions at even lower temperatures \cite{LBalentsPhys2011, TOjanenPRB2013, JMaPRB2015, SZhongPRL2016, CFangNP2016, QMaNP2017, FdJuanNC2017, GChangPRL2017, FFlickerarXiv2018}.
 
 In conclusion, the ARPES measurements on CoSi revealed the spectral signatures of two types of exotic chiral fermions, the spin-1 chiral fermion and the double Weyl fermion, by elucidating the unconventional multifold band crossings at the $\Gamma$ and $R$ points of the bulk BZ. The measurements also revealed the Fermi-arc surface states associated with these fermions. The present result establishes the existence of chiral topological fermions characterized by the Chern number larger than 1, and lays a foundation for studying unconventional physical properties related to the chiral fermions.

\begin{acknowledgments}
We thank K. Hori, K. Sugawara, C. X. Trang, D. Shiga, K. Tanaka, and S. Ideta  for their assistance in the ARPES measurements. We also thank Diamond Light Source for access to beamline I05 (proposal number S18839), KEK-PF for beamline BL2A (Proposal number: 2018S2-001 and 2016G-555), and UVSOR for beamline BL05U (Proposal number: 30-846).  This work was supported by MEXT of Japan (Innovative Area ``Topological Materials Science'' JP15H05853), JSPS (JSPS KAKENHI No: JP17H01139, JP26287071, JP25220708, JP16K13664, JP16K17725, and JP18H01160), and DFG (CRC1238 ``Control and Dynamics of Quantum Materials'', Project A04).
\end{acknowledgments}

\bibliographystyle{prsty}

\clearpage

{

\onecolumngrid
\begin{center}
{\large Supplemental Material for \\
``Observation of Chiral Fermions with a Large Topological Charge and Associated Fermi-Arc Surface States in CoSi''}

\vspace{0.3 cm}

Daichi Takane,$^1$ Zhiwei Wang,$^2$ Seigo Souma,$^{3,4}$ Kosuke Nakayama,$^1$ Takechika Nakamura,$^1$ Hikaru Oinuma,$^1$ Yuki Nakata,$^1$ Hideaki Iwasawa,$^5$ Cephise Cacho,$^5$ Timur Kim,$^5$ Kouji Horiba,$^6$ Hiroshi Kumigashira,$^{6,7}$ Takashi Takahashi,$^{1,3,4}$ Yoichi Ando,$^2$ and Takafumi Sato$^{1,3}$

{\footnotesize
$^1${\it Department of Physics, Tohoku University, Sendai 980-8578, Japan}
\newline
$^2${\it Physics Institute II, University of Cologne, 50937 K\"oln, Germany}
\newline
$^3${\it Center for Spintronics Research Network, Tohoku University, Sendai 980-8577, Japan}
\newline
$^4${\it WPI Research Center, Advanced Institute for Materials Research, Tohoku University, Sendai 980-8577, Japan}
\newline
$^5${\it Diamond Light Source, Harwell Science and Innovation Campus, Didcot, Oxfordshire OX11 0QX, UK}
\newline
$^6${\it Institute of Materials Structure Science, High Energy Accelerator Research Organization (KEK), Tsukuba, Ibaraki 305-0801, Japan}
\newline
$^7${\it Institute of Multidisciplinary Research for Advanced Materials (IMRAM), Tohoku University, Sendai 980-8577, Japan}
}

\end{center}



\renewcommand{\thefigure}{S\arabic{figure}}
\setcounter{figure}{0}

\vspace{0.5 cm}
\begin{center}
{\textbf {\large{S1. Experiments and band calculations}}}
\end{center}
\vspace{0.3 cm}

High-quality single crystals of CoSi were synthesized with a chemical vapor transport method by using I$_2$ as transport agent. High-purity Co powder (99.99\%) and Si powder (99.99\%) were sealed in an evacuated quartz tube, which was put in a two-zone tube furnace. The temperatures were set to 1000 $^{\circ}$C (source side) and 900 $^{\circ}$C (growth side), and kept for 10 days, after which single crystals (about 1 mm in size) with well-defined faces were obtained. ARPES measurements were performed with Scienta-Omicron R4000, SES2002, and MBS A-1 electron analyzers with energy-tunable synchrotron light at the beamline I05 in DIAMOND Light Source (DLS), BL-2A in Photon Factory (PF), KEK, and BL5U in UVSOR. We used linearly polarized light (horizontal polarization) of 130-240 eV in DLS, 260-630 eV in PF, and 56-72 eV in UVSOR. The energy and angular resolutions were set to be 10-150 meV and 0.2$^{\circ}$, respectively. Crystals were cleaved \textit{in-situ} in an ultrahigh vacuum better than 1$\times$10$^{-10}$ Torr along the (001) crystal plane. Sample was kept at $T$ = 25 K during ARPES measurements. CoSi single crystal is very hard to cleave and even the cleaved surface has several rough steps, as shown in a typical optical-microscope image in Fig. S1(a). When the light spot was on the rough surface, we did not observe any band dispersions. We have cleaved crystals more than 50 times in an UHV chamber, and in a very few cases, we have succeeded in cleaving and focusing the light beam on a small flat surface area, as highlighted by dashed red circle in Fig. S1(a). This flat area is as large as $\sim$200$\times$200 $\mu$m$^2$ which is larger than the beam-spot size. 

\begin{figure}[h]
\includegraphics[width=2.8 in]{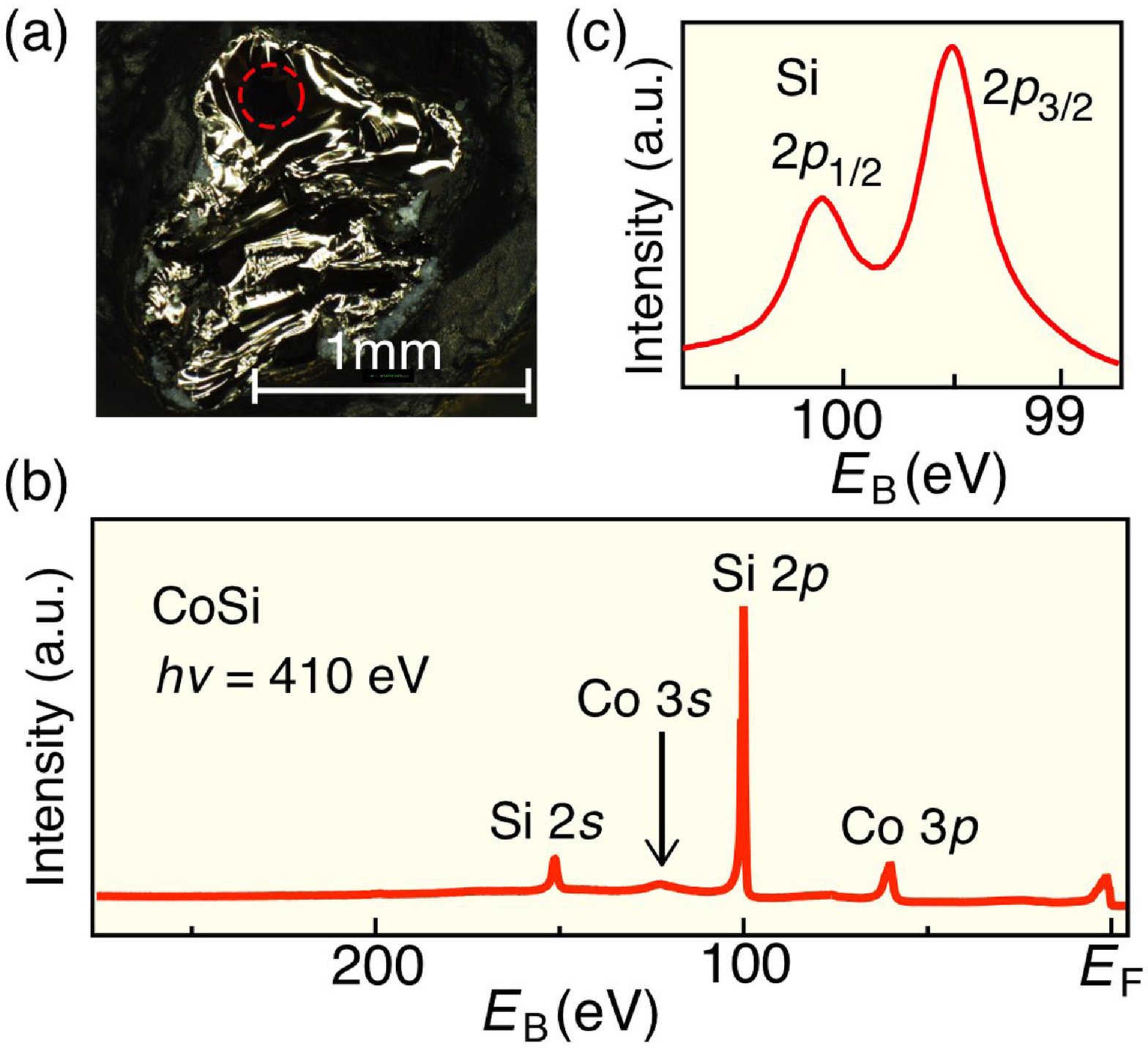}
\caption{(a) Optical-microscope image of cleaved CoSi crystal with a flat surface area indicated by red circle. (b) EDC of CoSi in a wide energy range measured with $h\nu$ = 410 eV. (c) Expansion of the Si2$p$ core-level.
}
\end{figure}

First-principles band-structure calculations were carried out using QUANTUM ESPRESSO code with generalized gradient approximation [37]. The plane-wave cutoff energy and the $k$-point mesh were set to be 30 Ry and 24 $\times$ 24 $\times$ 12, respectively.

Figure S1(b) displays the energy-distribution curve (EDC) in a wide energy range measured at photon energy ($h\nu$) of 410 eV. One can recognize several core-level peaks which are assigned to the Co (3$s$, 3$p$) and Si (2$s$, 2$p$) orbitals. The absence of additional features around the Si 2$p$ states as shown in Fig. S1(c) suggests no obvious oxidation, supportive of the clean surface. Besides the core levels, one can find a peak structure near $E_{\rm F}$ originating from the valence band composed of the Co 3$d$ and Si 3$p$ orbitals.

\vspace{0.5 cm}
\begin{center}
{\textbf {\large{S2. Analysis of the spin-1-chiral-fermion dispersion around the $\Gamma$ point}}}
\end{center}
\vspace{0.3 cm}

To discuss the band-crossing behavior of the spin-1 chiral fermions in CoSi in detail, we have carried out the spectral analysis around the $\Gamma$ point. It is well known that momentum-distribution curves (MDCs) and EDCs are suitable for mapping out the band dispersions of highly and weakly dispersive bands, respectively. We thus applied the MDC (EDC) analysis to the highly dispersive lower branch (nearly flat middle branch) of the spin-1-chiral-fermion bands. As seen in the MDCs along the $\Gamma$$M$ cut at $h\nu$ = 170 eV in Fig. S2(a), two peaks stemming from the lower branch are resolved at the binding energy ($E_{\rm B}$) of 0.4 eV. These peaks gradually move toward the $\Gamma$ point ($k_{x=y}$ = 0) on approaching $E_{\rm F}$ and eventually merge into a single peak slightly below $E_{\rm F}$. We numerically fitted the MDCs with two Lorentzians to extract the energy position of bands by assuming a linear dispersion. The middle branch is supposed to contribute little to the MDC shape around the $\Gamma$ point because of its flat feature. From the fittings, we have estimated the band top at the $\Gamma$ point to be at $E_{\rm B}$ = 20$\pm$20 meV. As shown in Fig. S2(b), we also estimated the location of middle branch at the $\Gamma$ point to be 5$\pm$5 meV from the EDCs divided by the Fermi-Dirac (FD) function at $T$ = 25 K convoluted with the resolution function. These values support the band degeneracy at the $\Gamma$ point as highlighted by the combined band dispersions in Fig. S2(c), although the possibility that a small spin-orbit-coupling (SOC)-induced gap may open is not finally excluded because of the finite experimental resolution. It is noted that the SOC produces two different band crossings near $E_{\rm F}$ separated from each other by $\sim$50 meV in the calculation [25], and this value is regarded as a magnitude of the spin-orbit gap. While it is difficult to experimentally distinguish the spin-orbit gap and the gap-like feature caused by the $k_{\rm z}$ broadening, the experimental uncertainty in estimating the energy position of lower and middle spin-1-chiral-fermion branches at the $\Gamma$ point sets the upper limit of the SOC-induced gap to be $\sim$40 meV at the $\Gamma$ point; this value corresponds to the largest energy interval between the lower and middle branches, located at 20$\pm$20 meV and 5$\pm$5 meV, respectively, when we take into account the experimental errors.

\begin{figure}[h]
\includegraphics[width=3.4 in]{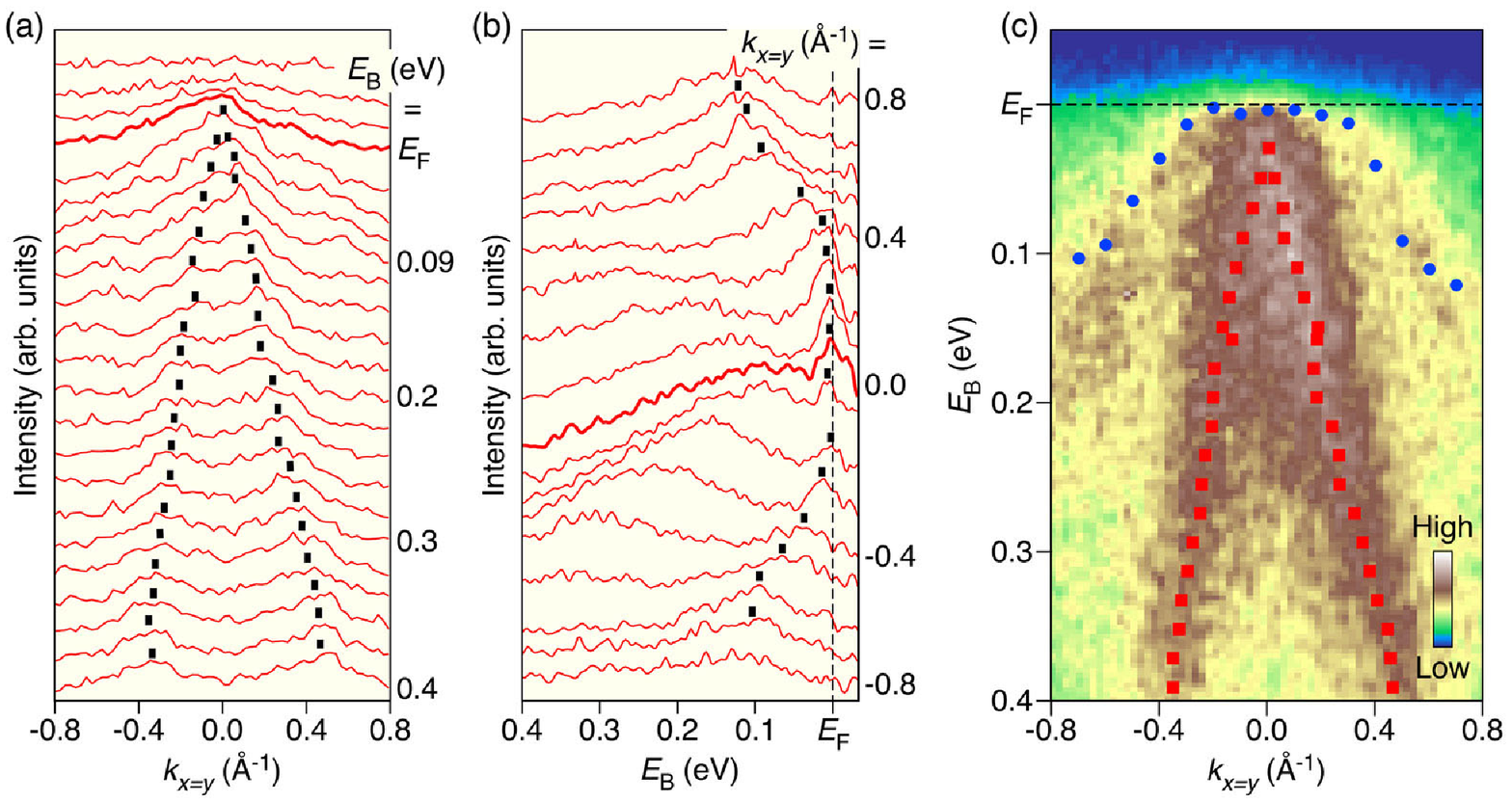}
\caption{(a) MDCs at $T$ = 25 K for several $E_{\rm B}$ slices along the ${\Gamma}M$ cut measured at $h\nu$ = 170 eV. Black dots indicate the peak position estimated by numerical fittings with two Lorentzians. (b) EDCs around the $\Gamma$ point divided by the FD function at 25 K convoluted with the resolution function. Black dots indicate the peak position of the middle branch. (c) Plot of ARPES intensity along the ${\Gamma}M$ cut together with the band dispersions estimated from the MDCs (red dots) and EDCs (blue dots).
}
\end{figure}

\vspace{0.5 cm}
\begin{center}
{\textbf {\large{S3. Double Weyl cone dispersion around the $R$ point}}}
\end{center}
\vspace{0.3 cm}

To discuss in detail the band dispersion related to the double Weyl fermions, we show in Fig. S3(a) and S3(b) the second-derivative intensity plot and the raw EDCs of CoSi measured along the $\Gamma$$RM$ cut in the bulk BZ. Measured \textbf{k} points for the $\Gamma$$R$ cut shown in Fig. S3(c) were accessed by simultaneously tuning the photon energy ($h\nu$  = 136.4-181.2 eV) and the sample angle. We also show in Figs. S3(d) and S3(e) similar data along the $\Gamma$$R$ cut, but obtained with higher-energy soft-x-ray (SX) photons of $h\nu$ = 337-421 eV [see Fig. S3(f)]. The overall structure of band dispersions obtained with the low- and high-energy photons are basically consistent with each other when the difference in the energy resolution between two measurements (20 and 150 meV) is considered.This confirms the validity of our experimental procedure to trace the $\Gamma$$R$ cut. In the EDCs [Figs. S3(b) and S3(e)], on can recognize two broad peaks at $E_{\rm B}$$\sim$0.4 eV and near $E_{\rm F}$ (black dots), respectively, midway between the $\Gamma$ and $R$ points. The former peak rapidly moves toward $E_{\rm F}$ on approaching the $\Gamma$ point, in agreement with the predicted dispersion of the spin-1 chiral fermions. As visible in the SX data of Figs. S3(d) and S3(e), these two bands gradually merge into a single band on approaching the $R$ point, consistent with the predicted dispersion of the double Weyl fermions. It is noted that in the low-photon energy data in Fig. S3(b) the splitting of lower double Weyl-cone branch is not so clear around the $R$ point due to the intensity suppression of the lowest double Weyl-cone branch. This behavior is in contrast to the SX data in Fig. S3(e) with the well visible lowest double Weyl-cone branch, likely due to the matrix-element effect.

\begin{figure}
\begin{center}
\includegraphics[width=4.5 in]{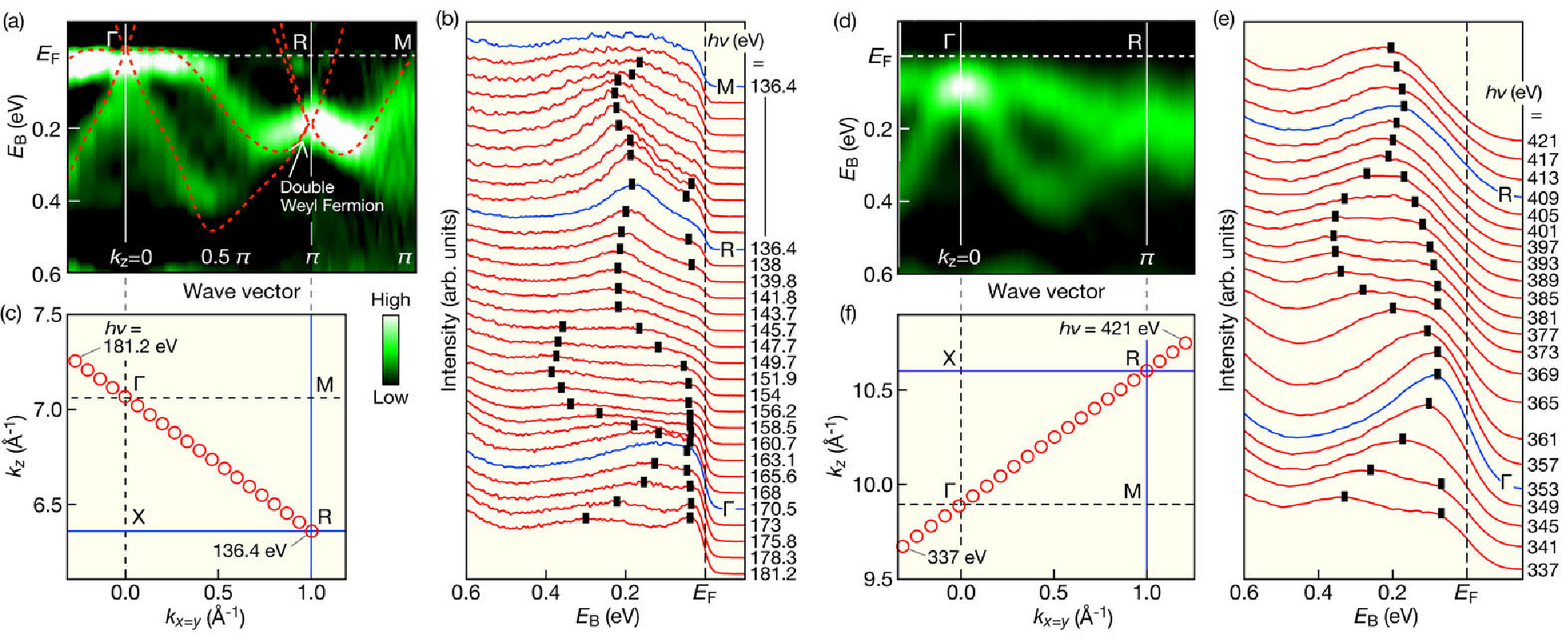}
\caption{(a), (b) Second-derivative intensity plot of EDCs and corresponding raw EDCs, measured along the $\Gamma$$RM$ cut of bulk BZ with  $h\nu$ = 136.4-181.2 eV. (c) Measured $k$ points for the $\Gamma$$R$ cut in the $k_{x=y}$-$k_z$ plane. (d)-(f) Same as (a)-(c) but measured along the $\Gamma$$R$ cut with SX photons of $h\nu$  = 337-421 eV.
}
\end{center}
\end{figure}

\vspace{0.5 cm}
\begin{center}
{\textbf {\large{S4. Absence of bulk Fermi surface around the $M$ point}}}
\end{center}
\vspace{0.3 cm}

To further substantiate our claim that only the novel fermions around the $\Gamma$ and $R$ points are relevant to the low-energy physics, it is important to experimentally confirm the absence of bulk Fermi surface in other $k$ regions. For this sake, we focused on the states around the $M$ point, since the second-derivative plot in Fig. 2(c) suggests the possibility for the existence of an energy band close to $E_{\rm F}$ in this momentum region. 
\begin{figure}
\includegraphics[width=4.5 in]{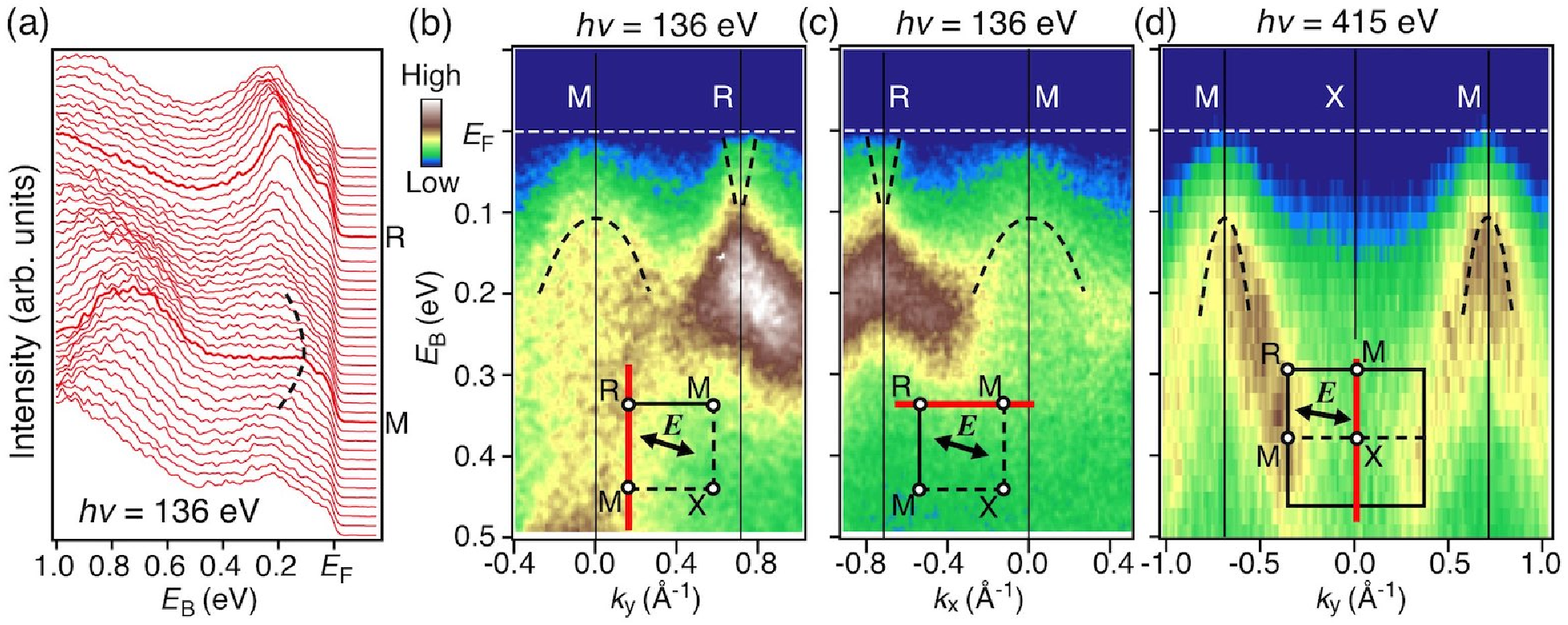}[b]
\caption{(a), (b) Raw EDCs and corresponding ARPES-intensity plot measured at $h\nu$ = 136 eV along the $MR$ cut. (c) Same as (b) but with 90$^{\circ}$-rotated polarization. (d) ARPES-intensity plot measured with SX photons of $h\nu$ = 415 eV along the $MX$ cut. Polarization of incident light is indicated by black arrow in the inset to (b)-(d). Dashed curves are a guide to the eyes to trace the band dispersion around the $M$ point.
}
\end{figure}
We show in Figs. S4(a) and S4(b) raw EDCs and the corresponding ARPES-intensity plot measured at $h\nu$ = 136 eV along the $MR$ cut. One can see a band dispersing toward $E_{\rm F}$ on approaching the $M$ point. However, this band does not cross $E_{\rm F}$ but is topped at $E_{\rm B}$$\sim$0.12 eV at the $M$ point. We confirmed the absence of the band-crossing around the $M$ point by rotating the direction of light polarization with respect to the $k$ cut by 90$^{\circ}$, as shown in Fig. S4(c). Moreover, we traced the $MX$ cut with SX photons of $h\nu$ = 415 eV, and confirmed no band crossings around the $M$ point, as highlighted in Fig. S4(d). These results suggest the absence of bulk Fermi surface around the M point.

\vspace{0.5 cm}
\begin{center}
{\textbf {\large{REFERENCES}}}
\end{center}
\vspace{0.3 cm}

\noindent
[25] P. Tang, Q. Zhou, and S.-C. Zhang, Phys. Rev. Lett. \textbf{119}, 206402 (2017).

\noindent
\hangindent = 14pt
\hangafter = 1
[37] P. Giannozzi, S. Baroni, N. Bonini, M. Calandra, R. Car, C. Cavazzoni, D. Ceresoli, G. L. Chiarotti, M. Cococcioni, I. Dabo, A. D. Corso, S. de Gironcoli, S. Fabris, G. Fratesi, R. Gebauer, U. Gerstmann, C. Gougoussis, A. Kokalj, M. Lazzeri, L. Martin-Samos, N. Marzari, F. Mauri, R. Mazzarello, S. Paolini, A. Pasquarello, L. Paulatto, C. Sbraccia, S. Scandolo, G. Sclauzero, A. P. Seitsonen, A. Smogunov, P. Umari, and R. M. Wentzcovitch, J. Phys.: Condens. Matter. \textbf{21}, 395502 (2015).

}

\end{document}